\input harvmac
\input epsf
\noblackbox
\overfullrule=0pt
\def\Title#1#2{\rightline{#1}\ifx\answ\bigans\nopagenumbers\pageno0\vskip1in
\else\pageno1\vskip.8in\fi \centerline{\titlefont #2}\vskip .5in}
 
scaled\magstep3 
 
scaled\magstep3 
 
scaled\magstep3 
 
\font\cmss=cmss10 \font\cmsss=cmss10 at 7pt
%

\newcount\figno
\figno=0
\def\fig#1#2#3{
\par\begingroup\parindent=0pt\leftskip=1cm\rightskip=1cm\parindent=0pt
\baselineskip=11pt \global\advance\figno by 1 \midinsert
\epsfxsize=#3 \centerline{\epsfbox{#2}} \vskip 12pt {\bf Fig.\
\the\figno: } #1\par
\endinsert\endgroup\par
}
\def\figlabel#1{\xdef#1{\the\figno}}
\def\encadremath#1{\vbox{\hrule\hbox{\vrule\kern8pt\vbox{\kern8pt
\hbox{$\displaystyle #1$}\kern8pt} \kern8pt\vrule}\hrule}}
\font\cmss=cmss10 \font\cmsss=cmss10 at 7pt

\def\IB{\relax\hbox{$\inbar\kern-.3em{\rm B}$}}
\def\IC{\relax\hbox{$\inbar\kern-.3em{\rm C}$}}
\def\IQ{\relax\hbox{$\inbar\kern-.3em{\rm Q}$}}
\def\ID{\relax\hbox{$\inbar\kern-.3em{\rm D}$}}
\def\IE{\relax\hbox{$\inbar\kern-.3em{\rm E}$}}
\def\IF{\relax\hbox{$\inbar\kern-.3em{\rm F}$}}
\def\IG{\relax\hbox{$\inbar\kern-.3em{\rm G}$}}
\def\IGa{\relax\hbox{${\rm I}\kern-.18em\Gamma$}}
\def\IH{\relax{\rm I\kern-.18em H}}
\def\IK{\relax{\rm I\kern-.18em K}}
\def\IL{\relax{\rm I\kern-.18em L}}
\def\IP{\relax{\rm I\kern-.18em P}}
\def\IR{\relax{\rm I\kern-.18em R}}
\def\Z{\relax\ifmmode\mathchoice
{\hbox{\cmss Z\kern-.4em Z}}{\hbox{\cmss Z\kern-.4em Z}}
{\lower.9pt\hbox{\cmsss Z\kern-.4em Z}} {\lower1.2pt\hbox{\cmsss
Z\kern-.4em Z}}\else{\cmss Z\kern-.4em Z}\fi}

\def\II{\relax{\rm I\kern-.18em I}}

\def\S{{\bf S}}

\def\p{\partial}

\lref\SchomerusVV{ V.~Schomerus, ``Rolling tachyons from Liouville
theory,'' arXiv:hep-th/0306026.
}

\lref\MooreZV{
G.~W.~Moore, M.~R.~Plesser and S.~Ramgoolam,
``Exact S matrix for 2-D string theory,''
Nucl.\ Phys.\ B {\bf 377}, 143 (1992)
[arXiv:hep-th/9111035].
}

\lref\TakayanagiSM{ T.~Takayanagi and N.~Toumbas, ``A matrix model
dual of type 0B string theory in two dimensions,'' JHEP {\bf
0307}, 064 (2003) [arXiv:hep-th/0307083].
}
\lref\DaCunhaFM{ B.~C.~Da Cunha and E.~J.~Martinec, ``Closed
string tachyon condensation and worldsheet inflation,''
arXiv:hep-th/0303087.
}
\lref\GrossAY{ D.~J.~Gross and N.~Miljkovic, ``A Nonperturbative
Solution of $D = 1$ String Theory,'' Phys.\ Lett.\ B {\bf 238},
217 (1990);
}
%
\lref\BrezinSS{ E.~Brezin, V.~A.~Kazakov and A.~B.~Zamolodchikov,
``Scaling Violation in a Field Theory of Closed Strings in One
Physical Dimension,'' Nucl.\ Phys.\ B {\bf 338}, 673 (1990);
}
%
\lref\GinspargAS{ P.~Ginsparg and J.~Zinn-Justin, ``2-D Gravity +
1-D Matter,'' Phys.\ Lett.\ B {\bf 240}, 333 (1990).
}
\lref\DasKA{ S.~R.~Das and A.~Jevicki, ``String Field Theory And
Physical Interpretation Of D = 1 Strings,'' Mod.\ Phys.\ Lett.\ A
{\bf 5}, 1639 (1990).
}
\lref\DouglasUP{ M.~R.~Douglas, I.~R.~Klebanov, D.~Kutasov,
J.~Maldacena, E.~Martinec and N.~Seiberg, ``A new hat for the c =
1 matrix model,'' arXiv:hep-th/0307195.
}
\lref\StromingerFN{ A.~Strominger and T.~Takayanagi, ``Correlators
in timelike bulk Liouville theory,'' arXiv:hep-th/0303221.
}

\lref\kkk{ V.~Kazakov, I.~K.~Kostov and D.~Kutasov, `A matrix
model for the two-dimensional black hole,'' Nucl.\ Phys.\ B {\bf
622}, 141 (2002) [arXiv:hep-th/0101011].
}
\lref\GrossUB{ D.~J.~Gross and I.~R.~Klebanov, ``One-Dimensional
String Theory On A Circle,'' Nucl.\ Phys.\ B {\bf 344}, 475
(1990).
}
\lref\sen{ A.~Sen, ``Rolling tachyon,'' JHEP {\bf 0204}, 048
(2002) [arXiv:hep-th/0203211].
}
\lref\mgas{ M.~Gutperle and A.~Strominger, ``Spacelike branes,''
JHEP {\bf 0204}, 018 (2002) [arXiv:hep-th/0202210].
}
\lref\MinicRK{ D.~Minic, J.~Polchinski and Z.~Yang, ``Translation
Invariant Backgrounds In (1+1)-Dimensional String Theory,'' Nucl.\
Phys.\ B {\bf 369}, 324 (1992).
}
\lref\MartinecKA{ E.~J.~Martinec, ``The annular report on
non-critical string theory,'' arXiv:hep-th/0305148.
}
\lref\PolchinskiMB{
J.~Polchinski,
``What is string theory?,''
arXiv:hep-th/9411028.
}

\lref\kms{ I.~R.~Klebanov, J.~Maldacena and N.~Seiberg, ``D-brane
decay in two-dimensional string theory,'' JHEP {\bf 0307}, 045
(2003) [arXiv:hep-th/0305159].
}
\lref\GinspargIS{
P.~Ginsparg and G.~W.~Moore,
``Lectures On 2-D Gravity And 2-D String Theory,''
arXiv:hep-th/9304011.
}
\lref\hv{ J.~McGreevy and H.~Verlinde, ``Strings from tachyons:
The c = 1 matrix reloaded,'' arXiv:hep-th/0304224.
}
\lref\msy{ A.~Maloney, A.~Strominger and X.~Yin, ``S-brane
thermodynamics,'' arXiv:hep-th/0302146.
}
\lref\BrezinRB{ E.~Brezin and V.~A.~Kazakov, ``Exactly Solvable
Field Theories Of Closed Strings,'' Phys.\ Lett.\ B {\bf 236}, 144
(1990).
}
\lref\GrossVS{ D.~J.~Gross and A.~A.~Migdal, ``Nonperturbative
Two-Dimensional Quantum Gravity,'' Phys.\ Rev.\ Lett.\  {\bf 64},
127 (1990).
}
\lref\DouglasVE{ M.~R.~Douglas and S.~H.~Shenker, ``Strings In
Less Than One-Dimension,'' Nucl.\ Phys.\ B {\bf 335}, 635 (1990).
}
\lref\PolchinskiUQ{ J.~Polchinski, ``Classical Limit Of
(1+1)-Dimensional String Theory,'' Nucl.\ Phys.\ B {\bf 362}, 125
(1991).
} \lref\KlebanovQA{ I.~R.~Klebanov, ``String theory in
two-dimensions,'' arXiv:hep-th/9108019.
}

\lref\NatsuumeSP{ M.~Natsuume and J.~Polchinski, ``Gravitational
Scattering In The C = 1 Matrix Model,'' Nucl.\ Phys.\ B {\bf 424},
137 (1994) [arXiv:hep-th/9402156].
}

\lref\KarczmarekXM{ J.~L.~Karczmarek, H.~Liu, J.~Maldacena and
A.~Strominger, ``UV finite brane decay,'' arXiv:hep-th/0306132.
}

\lref\GaiottoRM{ D.~Gaiotto, N.~Itzhaki and L.~Rastelli, ``Closed
strings as imaginary D-branes,'' arXiv:hep-th/0304192.
}

\lref\PolchinskiJP{ J.~Polchinski, ``On the nonperturbative
consistency of d = 2 string theory,'' Phys.\ Rev.\ Lett.\  {\bf
74}, 638 (1995) [arXiv:hep-th/9409168].
}

\lref\AlexandrovCM{
S.~Alexandrov and V.~Kazakov,
``Correlators in 2D string theory with vortex condensation,''
Nucl.\ Phys.\ B {\bf 610}, 77 (2001)
[arXiv:hep-th/0104094].
}

\lref\AlexandrovFH{
S.~Y.~Alexandrov, V.~A.~Kazakov and I.~K.~Kostov,
``Time-dependent backgrounds of 2D string theory,''
Nucl.\ Phys.\ B {\bf 640}, 119 (2002)
[arXiv:hep-th/0205079].
}

\lref\AlexandrovPZ{
S.~Y.~Alexandrov and V.~A.~Kazakov,
``Thermodynamics of 2D string theory,''
JHEP {\bf 0301}, 078 (2003)
[arXiv:hep-th/0210251].
}

\lref\AlexandrovQK{
S.~Y.~Alexandrov, V.~A.~Kazakov and I.~K.~Kostov,
``2D string theory as normal matrix model,''
Nucl.\ Phys.\ B {\bf 667}, 90 (2003)
[arXiv:hep-th/0302106].
}

%
\Title{\vbox{\baselineskip12pt \hbox{hep-th/0309138}}} {\vbox{
\centerline {Matrix Cosmology}}} \centerline{Joanna L. Karczmarek
  and  Andrew Strominger} 
\vskip.1in {\it Jefferson Physical Laboratory, Harvard
University, Cambridge, MA 02138}

\vskip.1in \centerline{\bf Abstract}
Exact time-dependent solutions of $c=1$ string theory are described
using the free fermion formulation. One such class of solutions describes
draining of the Fermi sea and has a spacetime interpretation
as closed string tachyon condensation. A second class of solutions,
corresponding to droplets of Fermi liquid orbiting in phase space,
describes closed cosmologies which bounce through singularities.

 \Date{}

\listtoc\writetoc
%
\newsec{Introduction}

The discovery of the double scaling limit of $c\le 1$ matrix
models \refs{\DouglasVE \GrossVS \BrezinRB \GrossAY \BrezinSS
-\GinspargAS} provided a beautiful and nonperturbative solution to bosonic
string theory in two spacetime dimensions \MooreZV. However these 
developments seemed not to fully find their place within the 
modern framework of string duality. Recently, the situation has 
changed with the elegant reinterpretation of these models as a soluble 
example of bulk-boundary duality \refs{\hv \MartinecKA -\kms} for  
type 0 strings \refs{\TakayanagiSM, \DouglasUP}.
With this new perspective  it is
natural to try to use these models to address various outstanding
puzzles in nonperturbative string theory. In this paper we will
look at {\it time-dependent} solutions of $c=1$ string theory.

  The mathematically simplest formulation of the $c=1$ theory is as a
theory of free fermions in an inverted harmonic oscillator
potential. In this setup the ground state is usually described as
having the Fermi sea filled up to  energy  $-\mu$, where $\mu$ is related
to the string coupling.  The point of view which we shall adopt in
this paper is that the full nonperturbative theory contains $all$
states of any number of free fermions in the inverted harmonic potential.
These states may or may not have a weakly coupled description as a
smooth $D=2$ spacetime geometry with tachyons.  However, we find
several examples in which such weakly coupled descriptions can be
derived for arbitrarily large regions of spacetime, while
other regions contain singularities which are resolved in the free
fermion description. Our analysis relies heavily on the
tachyon-fermion dictionary given in
\refs{\PolchinskiUQ,\DasKA}. One class of examples involves
tachyon condensation along a spacelike
\foot {The $g_s \to \infty$ limit of our solution first appeared in
\MinicRK.} or null
hypersurface.\foot{This involves a non-normalizable perturbation,
and hence does not ruin the stability of the usual ground state.}
The spacelike case at weak coupling corresponds to a worldsheet
with a timelike bulk Liouville factor, while the null case
involves (an analytic continuation of) the Sine-Gordon Liouville
theory. Both have exact descriptions in the free fermion picture
as a draining of the Fermi sea. A second class of solutions describes closed,
ekpyrotic cosmologies which begin at weak coupling and bounce off
a strong coupling region into a second weakly-coupled region. This
is described as a finite-size but macroscopic droplet of the Fermi
sea orbiting in phase space.

This paper is organized as follows. Section 2 briefly reviews the
$c=1$ matrix model and fixes our notation. Section 3 describes two
classes of semiclassical solutions of the free Fermi theory,
characterized as a moving liquid in phase space. It is then shown,
using the dictionary of \PolchinskiUQ, that these correspond
to tachyon condensation along null or spacelike slices. We further
go beyond the semiclassical limit and construct exact quantum
states in the free fermion picture corresponding to tachyon
condensation. Section 4 contains a closed cosmology with a large
weakly coupled region. Section 5 considers a more general class of
cosmologies which correspond to arbitrarily shaped droplets of
Fermi liquid orbiting in phase space. This project arose as a
search for a large $N$ decoupling limit of sD-branes, and in the last
section we describe how the Fermi droplets are a $c=1$ version of
such, pointing out also potential generalizations to higher
dimensions.

After initial submission of this paper, we became aware of 
work by S. Alexandrov, V. Kazakov and I. Kostov \AlexandrovFH, 
which overlaps with our paper. See also
\refs{\AlexandrovCM \AlexandrovPZ - \AlexandrovQK}
for related interesting results.

\newsec{Lightning review of the $c=1$ matrix model}

In the free fermion formulation the $c=1$ matrix model is described
by the hamiltonian\foot{
We set $\alpha' = 1$ and roughly follow the notation of 
\PolchinskiMB.}
 \eqn\ffr{{\cal H}=\half\int dx \bigl( \p_x
\Psi^\dagger \p_x\Psi-x^2\Psi^\dagger\Psi \bigr)~,} where
\eqn\ccr{\bigl\{ \Psi^\dagger(x),\Psi(x')\bigr\}= \delta(x-x')~.}
The usual ground state $|\mu\rangle$ is defined by filling the
Fermi sea up to an energy $E=-\mu$ : \eqn\ghj{b^\dagger_E
|\mu\rangle=0~,~~~~~~E <-\mu~,~~~~~~~~~~~~~~~b_E
|\mu\rangle=0~,~~~~~~ E>-\mu~,} where $b^\dagger_E$ creates a
fermion of energy $E$. This requires an infinite number of
fermions since the energy in the inverted harmonic oscillator is
unbounded from below.

For part of our discussion we  will be expanding about the
classical point particle limit in which the phase space volume per
fermion goes to zero and the Fermi sea can be treated as a
continuous liquid in phase space \PolchinskiUQ. The Fermi sea and
its boundary the Fermi surface move along classical trajectories
-- hyperbolas in phase space -- under time evolution. The ground
state corresponds to filling the phase space region bounded by the
classical trajectory with $E=-\mu$ as depicted in figure 1(a). Since
this boundary is invariant under time evolution, the ground state
is static. We will be interested in more general time dependent
states which may be semiclassically described by filling a region
of phase space at $t=0$ and then evolving it along the classical
trajectories.

The description of this system as a $c=1$ noncritical string
theory contains a tachyon field $T$ propagating in two spacetime
dimensions. This field is a bosonization of $\Psi$ describing
fluctuations in the Fermi surface of the ground state
$|\mu\rangle$. At any moment of time the value of the momentum $p$
on the Fermi surface is a double-valued function of $x$ whose
values we denote $p_\pm(x)$. In the ground state,
\eqn\ook{p_\pm=\pm \sqrt{x^2-g_s^{-1}}~,} where the distance of
closest approach to the origin is the inverse square root of the
string coupling, $g_s = 1/2\mu$. 
More generally, the classical equations of motion
\eqn\fdc{\dot p= x~,~~~~\dot x= p} imply \eqn\EoM{
\partial_t p_\pm = x - (\partial_x p_\pm) p_\pm~.
} Following \PolchinskiUQ, $p_\pm$ may be  written in terms of a
collective field $S$ and its canonical conjugate $\pi_S$:
\eqn\ppm{ p_\pm = \pm e^{-q} - \sqrt{\pi}e^q (\pi_S \mp \partial_q
S)~, } where $x = -e^{-q}$.  To match \ook, $S$ at large negative
$q$ must be \eqn\S{ S(q) = -{q \over 2 \sqrt{\pi} g_s}+ const~. }
The collective field $S$ is related to the tachyon $T$ in the
Liouville theory by \foot{
In work subsequent to \PolchinskiUQ, a distinction is made betweed
the string theory field $S$ and the matrix model collective field
$\bar S$ \NatsuumeSP.  For our purposes, this distinction is not 
important, and
we will not stress it.  $\bar S$ and $S$ are related by a nonlocal
transform which can be represented as a function of $d/dq$ acting
on $\bar S$.  For the specific forms of $S(q)$ which we will consider,
this will reduce to either a constant multiplicative factor, or
an additive constant.}
\eqn\ts{T\propto e^ {2 q } S~.}  This
correspondence is meaningful only for $x\rightarrow -\infty$,
where the worldsheet theory is weakly coupled. With $\alpha' = 1$,
the coordinates $(q,t)$ are related to the Liouville coordinates
$(X^1, X^0)$ by $X^0 = t$ and $X^1 = q + const$. Following
\PolchinskiUQ\foot{Except for the constant shift allowed between
$X^1$ and $q$ which has been chosen to simplify the expression.}
we choose the fudge factor in \ts\ and \S\ so that the tachyon
field corresponding to \S\ is \eqn\T{ T = -(\hat\mu X^1)e^{ 2 X^1}~.
} Including this field leads to the closed string worldsheet
action \eqn\action{ {1 \over {4\pi}} \int d^2\ \sigma \sqrt{\hat
g} \Big\{ \hat g^{ab} \eta_{\mu\nu}
\partial_a X^\mu \partial_b X^\nu + 2  X^1 \hat R - \hat \mu X^1
e^{2 X^1}\bigr\}~,} which is the $c=25$ Liouville theory supplemented
by a timelike boson.

A small incoming tachyon fluctuation can be described as a bump in
the Fermi sea which makes its way around the hyperbola. Using this
picture and the relation \ppm\ the tachyon S-matrix has been
computed from the free fermion picture and compared to 
results from standard
worldsheet methods based on \action. Agreement has been found -- a
review appears in \refs{\KlebanovQA,\GinspargIS}.

\newsec{Tachyon condensation and
draining the Fermi sea}

In this section, we will describe a simple time-dependent solution
of the free fermion model \ffr. Using the dictionary \ppm, we will
see that the deviation of the Fermi sea from  the ground state
\ook\ is too large for the solution to be interpreted as
weak-field tachyon scattering. Instead we will argue that it
corresponds to closed string tachyon condensation.
\subsec{Lightlike condensation}

A solution of the equations of motion for the Fermi surface is
given by\foot{For a Type 0 NS solution we should add a second
filled region at positive $x$. While the existence of a
nonperturbative Type 0 interpretation is important for our
discussion, the details are not, and we shall for the sake of
brevity suppress all discussion of the second region.}
\eqn\moving{ (x + p + 2\lambda_+ e^t)(x - p + 2\lambda_- e^{-t}) =
g_s^{-1}~,} where $\lambda_+, \lambda_-$ are arbitrary
non-negative constants. This is a $\it moving$ hyperbola centered
at $(x,p)= (-\lambda_+ e^t - \lambda_- e^{-t}, -\lambda_+ e^t +
\lambda_- e^{-t})$ rather than the origin.

To understand what \moving\  represents, let's consider some
special cases. Take $\lambda_+ = \lambda_-=\lambda$; at $t=0$
the filled portion of the Fermi sea is displaced by an amount
$-2\lambda$ in the $x$ direction (see figure 1(c)).  Under time
evolution fermions come in from $x=-\infty$, up to the point given
by $x=-2\lambda-1/\sqrt{g_s}$ and then back to $x=-\infty$.
Another interesting special case is $\lambda_-=0$. In the infinite
past the Fermi sea is essentially static and filled up to the
energy $g_s^{-1}$ below the top of the potential. It then decays
away (rolls down the potential) and all fermions move out to
$x=-\infty$ in the infinite future. \fig{(a) The usual ground
state has the Fermi sea filled up to the left constant energy
surface $E=-1/g_s$ denoted by the bold hyperbola. The arrowed
lines indicate classical trajectories. (b-d) At any fixed time the
general solution \moving\ is a displaced hyperbola, each point of
which  which is dragged along the classical
trajectories.}{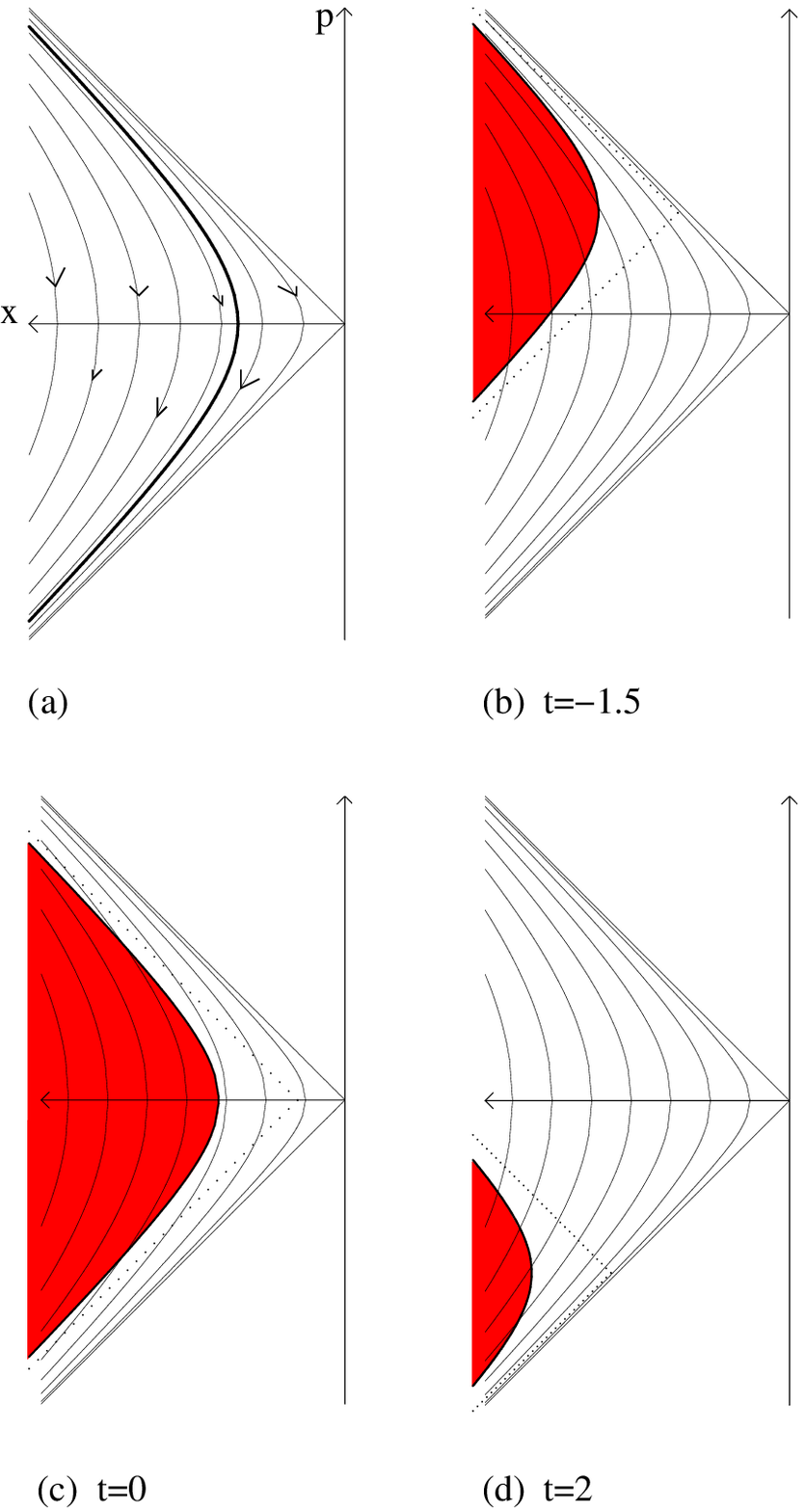}{0.0truein} Using the dictionary \ppm
in the asymptotic region of large $-x=e^{-q}$, the solution
\moving\ corresponds to \eqn\S{ \sqrt{\pi} S = - {1 \over 2 g_s }
q + \lambda_+ e^{t-q} + \lambda_- e^{-t-q} + o(e^{-2q})~, } up to
a constant. According to \ts, the tachyon field corresponding to
\S\ is 
\foot{We suppress the effect of the 
leg-pole factors, which can be accounted for 
by a nonlocal transform, as in equation (20) of \NatsuumeSP.  
Here, the effect of this transform is just a multiplicative factor,
which can be absorbed by a rescaling of the couplings and a shift in $X^1$. 
However, the kernel K has a pole for the precise
solution we are considering, resulting in a
divergence which must be regularized and renormalized.  
One regulator we can use is to
slightly change the interaction term, moving its euclidian momentum 
away from the pole in the kernel, while keeping
the resulting perturbation marginal.  }
 \eqn\T{ T = -\hat \mu  X^1 e^{2 X^1} + \hat \lambda_+
e^{X^1+X^0} + \hat \lambda_- e^{X^1-X^0}~, } where $\hat
\lambda_\pm$ are proportional to $\lambda_\pm$. Including this
field leads to the closed string worldsheet action \eqn\actionB{
{1 \over {4\pi}} \int d^2\ \sigma \sqrt{\hat g} \Big\{ 2  X^1 \hat
R - \hat \mu X^1 e^{2 X^1} + \hat \lambda_+ e^{X^1+X^0} + \hat
\lambda_- e^{X^1-X^0} \Big\}~. } 
Analytic continuation $X^0 \rightarrow -iX^2$ of this action
gives the Sine-Gordon Liouville action. 
The first
interaction term corresponds to the standard Liouville potential
which is a timelike ``wall'' at $X^1 \sim -\ln \hat \mu$. The
second and third interaction terms (which are weight (1,1)
operators) are also exponential potential walls, albeit less steep
than the standard one. These walls move with time, effectively
cutting off the universe at a time-varying distance. They are
located roughly along the ingoing and outgoing null trajectories
at $X^1\pm X^0\sim -\ln \hat \lambda_\pm$. Hence at large positive
(negative) $X^0$, the wall moves outwards (inwards) with
essentially the speed of light. The situation is depicted in
figure 2. \fig{Penrose diagram for lightlike tachyon condensation.
The jagged lines indicate the locations of the past, future and
timelike Liouville walls.}{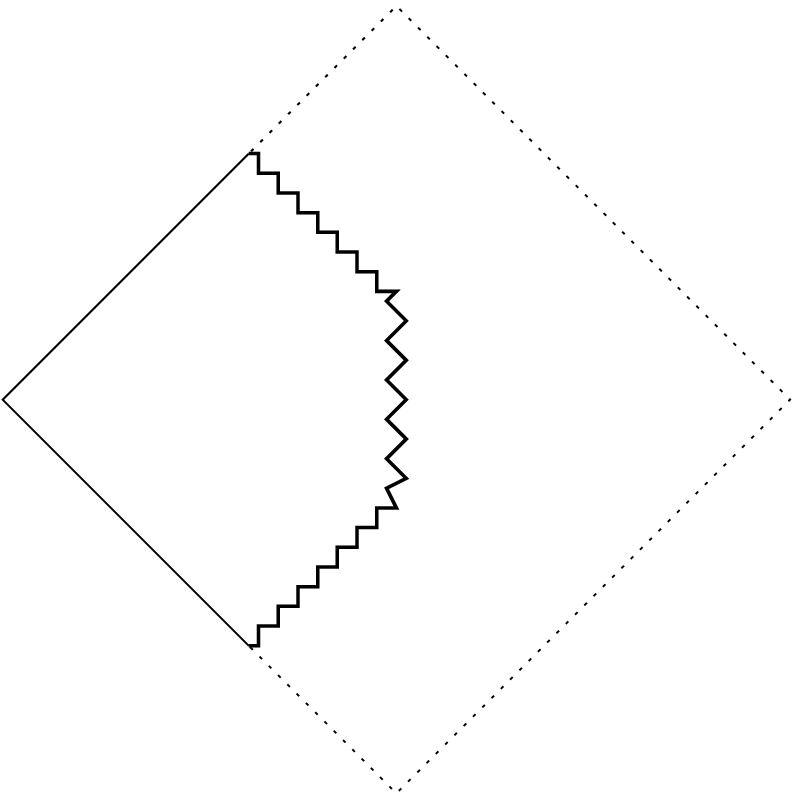}{0.0truein}

Any observer moving along a timelike trajectory will eventually
move into a region where the tachyon field becomes arbitrarily
large. Therefore this solution is a form of tachyon condensation.
An exact expression for the quantum state corresponding to this solutions
will be given in subsection 3.3.

A limit can be considered in which the first interaction term in
\actionB\ can be eliminated by setting $\hat \mu$ (and hence
$g_s^{-1}$) to zero. This corresponds to a Fermi surface which is
not a smooth moving hyperbola, but has a corner which itself moves
along a hyperbola. The strong coupling region is now hidden by the
null ingoing and outgoing potential walls.

A further limit may be taken in which all but the last term are
set to zero.  In the far past, the remaining term vanishes and we
have the linear dilaton vacuum with no Liouville wall. At
any finite time a wall is present and moving out to spatial
infinity at the speed of light. In the far future the tachyon has
condensed and the fermions are all gone. This corresponds to
draining of the Fermi sea. However, this limiting situation is
singular in the tachyon description, since at $t=-\infty$ the
strong coupling region is exposed.

\subsec{Spacelike condensation} In this section, we present
another method of draining the Fermi sea, which can also be
interpreted as tachyon condensation, but on a spacelike surface.

The general solution is
 \eqn\wedge{ \Big (\sinh(t-T_1) x -
\cosh(t-T_1) p \Big ) \Big (\sinh(t-T_2) x - \cosh(t-T_2) p \Big )
=
 - {1\over 2 g_s} \cosh(T_2-T_1)
~.} The $g_s \to \infty$ limit of this solution appeared first in
\MinicRK.\foot{ We thank J. Polchinski for drawing this to our
attention.} We first consider this solution in the $T_1 \rightarrow
-\infty$, $g_s \to \infty$ limit \eqn\halfwedge{ e^t \Big (x - p
\Big ) \Big (\sinh(t) x - \cosh(t) p \Big ) = 0~. } This
corresponds to a ``wedge'' of the Fermi sea. At $t=-\infty$, it is
the entire quadrant $p+x>0, ~~p-x<0$, which is the linear dilaton
vacuum without a tachyon wall. As time passes, the upper edge of
the Fermi sea remains a straight line, but rotates
counterclockwise in such a fashion that it has slope $\tanh(t)$.
At $t =+\infty$, the upper and lower edges coincide, and the Fermi
sea has disappeared. This is depicted in figure 3. At early times
the tachyon field is small and we can use \ppm\ and \ts\ to
compute its value. We find for $X^0\to - \infty$ \eqn\iio{T\sim
e^{2X^0}.} Unlike the previous case, here we find a tachyon field
which is constant along spacelike (rather than null) slices. \iio\
is the exactly marginal timelike bulk Liouville interaction
considered in \refs{\DaCunhaFM \StromingerFN -\SchomerusVV}.
\fig{Tachyon condensation on a spacelike surface in the linear
dilaton vacuum is represented by the upper edge of the Fermi sea
undergoing counterclockwise rotation. In the infinite future it
meets the lower edge and the fermions have all disappeared.
}{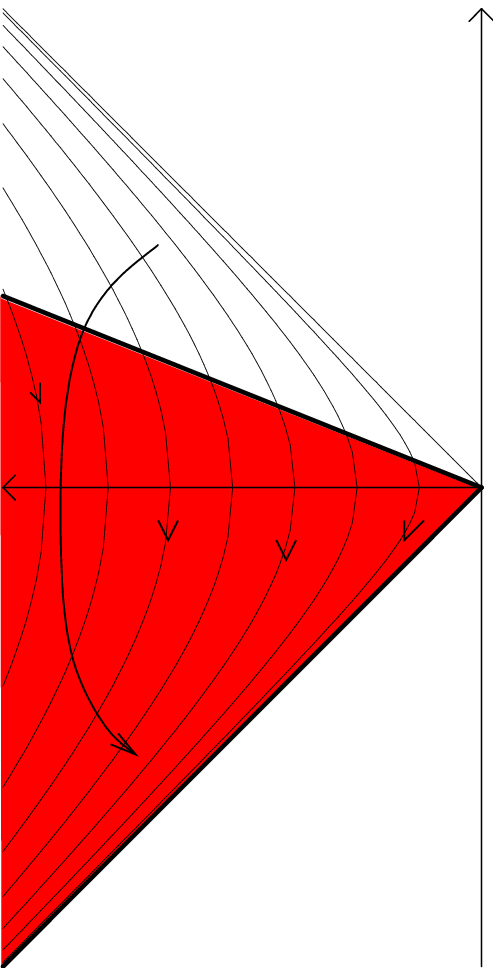}{0.0truein}

The above case of pure linear dilaton is singular because there is
no tachyon wall shielding the strong coupling region. This wall is
reinstated by taking $g_s \neq 0$, so that \eqn\halfwedge{ e^t
\Big (x - p \Big ) \Big (\sinh(t) x - \cosh(t) p \Big ) =
 - {1\over 2 g_s}~.
} This corresponds to the worldsheet action \eqn\actionC{ {1 \over
{4\pi}} \int d^2\ \sigma \sqrt{\hat g} \Big\{ \hat g^{ab}
\eta_{\mu\nu} \partial_a X^\mu \partial_b X^\nu + 2  X^1 \hat R -
\hat \mu X^1 e^{2 X^1} + \kappa e^{2 X^0} \bigr\}~, } where
$\kappa$ is a constant that could be absorbed by a shift of $X^0$.
Although this expression -- and indeed the entire dictionary
between free fermions and tachyons -- was derived for weak tachyon
fields, it is notable that \actionC\ is in fact the direct sum of
two noninteracting CFTs, and therefore provides an exact string
tree-level solution.

The general solution \wedge\ corresponds to a very narrow wedge of
Fermi sea coming in from $-\infty$, opening up to the standard
hyperbola at $t = T_1$ (we take $T_1<T_2$ for concreteness) and
then closing back down at $t = T_2$. Solving for $p_\pm$, and
using equations \ppm\ and \S, the tachyon field is \eqn\T{ T =
-\hat\mu  X^1 e^{2 X^1} + \kappa (1-f(X^0))~, } where \eqn\f{ f(t)
= {\sinh(T_2 - T_1) \over 2 \cosh(t-T_1) \cosh(t-T_2)}~. } As
before, the new term in the tachyon field has no spatial
dependence.  For $T_1 < X^0 < T_2$ it vanishes, and we recover
regular Liouville theory \action.  Let's take $T_2-T_1 \gg 1$ so
that the Liouville universe exists for an appreciable amount of
time. Before and after this (arbitrarily long) period, the
coupling becomes strong everywhere, and smooth spacetime
disappears.  This presumably corresponds to a process in which the
tachyon is condensed in both the far future and the far past.
Actually, equation \T\ is meaningful only for small $X^0$
(for $T_1 < X^0 < T_2$), where the expression is \eqn\T{ T = -\hat
\mu  X^1 e^{2 X^1} + \kappa \Big(e^{2(X^0 - T_2)} +e^{-2(X^0 -
T_1)} \Big)~. }

\subsec{Exact quantum solutions}

In this section we give exact quantum solutions with the
prescribed semiclassical behavior. First we consider the general
spacelike tachyon condensation of equation \wedge. We begin by
introducing the dilation operator \eqn\dlo{D_\alpha =e^{{i 
\alpha \over 2}(\hat p \hat x + \hat x \hat p)}} 
which stretches $x$ and shrinks $p$ \eqn\jhl{ D_\alpha \hat  x
D_{-\alpha}=e^{\alpha} \hat  x~,~~~~~~ D_\alpha \hat p D_{-\alpha }
=e^{-\alpha}\hat  p~.} The hat on $\hat x$ and $\hat p$ indicates they
are operators. Now consider the state at time $t=0$
\eqn\rfs{|\alpha,\mu\rangle = D_\alpha |\mu \rangle~,} where $|\mu
\rangle$ is the ground state \ghj, and the action $D_\alpha$ on a
second quantized state is 
induced from its action on the first quantized wave functions,
and can be written as
\eqn\dfv
{D_{\alpha} = e^{-{\alpha \over 2} \int \big (dx ~\Psi^\dagger
(x \partial_x \Psi ) - ( x \partial_x \Psi^\dagger
) \Psi  \big )}
~.}
Since $D_\alpha$ does not
commute with the Hamiltonian, $|\alpha,\mu\rangle $ is a time
dependent state. At time $t=0$, in the semiclassical
approximation, the Fermi surface is described by the equation
\eqn\youi{e^{2\alpha}p^2-e^{-2\alpha}x^2=-{1 \over g_s}~.} Taking
$e^{-2\alpha} = \tanh{T_2-T_1 \over 2}$  and rescaling $g_s$ by
$2 \sinh {T_2-T_1 \over 2}$, we get the general solution \wedge\ at time
$t=\half(T_1+T_2)$. Therefore, evolving $|\alpha,\mu\rangle $ with
$H$ we get an exact quantum state whose Fermi surface evolves
under time approximately according to \wedge. Fluctuations in this
Fermi surface are suppressed, and the semiclassical limit
approached, for example by taking $\mu$ to be very large.

The exact quantum state corresponding to \moving\ is even easier
to find -- it can be created by acting on $| \mu \rangle $ with
the shift operator of the form $e^{-i\lambda_-(\hat p+\hat x) -
i\lambda_+(\hat p-\hat x)}$. 

Expansion around the static ground state of the matrix model has
provided a powerful method of computing CFT correlators in the
usual Liouville theory. Here we have presented simple exact
quantum states related to the timelike bulk Liouville theory and
the Sine-Gordon Liouville theory. Perhaps the corresponding matrix
states will be of use in computing correlators of these CFTs.

\subsec{Stability}

The reader may be puzzled by the fact that the closed string
tachyons in the Liouville theory are often referred to as
massless, and thus can not condense in the usual  sense. The
possibility of tachyon condensation seems to be at odds with the
stability of the $c=1$ matrix model. However the interactions
$e^{X^1\pm X^0}$ and $e^{\pm 2 X^0}$ are not normalizable
fluctuations of the tachyon field. Therefore the state
$|\mu\rangle$ does not decay via such fluctuations and there is no
problem of stability.

\newsec{Closed cosmologies}
\subsec{Orbiting Fermi droplet}In this subsection we construct a
matrix cosmology from $N<\infty$ fermions in the $c=1$ matrix
model. The construction will proceed in three steps.

In the first step we consider a state, involving an infinite
number of fermions, in which every single-fermion eigenstate with
energy {\it greater} than $-\mu$ is filled. Due to the Pauli
exclusion principle, this state obeys  \eqn\ghsj{b^\dagger_E
|\tilde \mu\rangle =0~,~~~~~~E >-\mu~,~~~~~~~~~~~~~~~b_E |\tilde
\mu\rangle =0~,~~~~~~ E <-\mu~,} where $b^\dagger_E$ creates a
fermion of energy $E$. Redefining \eqn\dlo{b^\dagger_E \to
a_E~,~~~~~~b_E \to a^\dagger_E~,} we find that $|\tilde
\mu\rangle$ obeys \eqn\ppl{ a_E|\tilde
\mu\rangle=0~,~~~~~~~~E>-\mu~,~~~~~~~a^\dagger_E|\tilde
\mu\rangle=0~,~~~~E<-\mu~,} which is just the usual condition for
a state whose Fermi sea is filled up to energy $\mu$. Hence the
state \ghj\ is equivalent to the usual matrix model ground state
(up to a shift and sign reversal of the energy implied by \dlo).
Excitations of this state can be described by a propagating
tachyon. This is just the usual particle-hole duality.

Next we consider the state \eqn\ghcj{\eqalign{b^\dagger_E
|\mu_1~,\mu_2\rangle =0~,&~~~-\mu_2<E <-\mu_1~,\cr ~~~~~b_E
|\mu_1~,\mu_2\rangle =0~,&~~~ E<-\mu_2,~~~E>
-\mu_1~,~~~~~~~~\mu_1<\mu_2.}} This state has two Fermi surfaces,
one at $\mu_1$ and the other at $\mu_2$. Small perturbations of
the Fermi surfaces are classically decoupled. This means that the
tree level tachyon S-matrix of this system, which in turn
determines the full perturbative S-matrix via unitarity, is
equivalent to two decoupled $c=1$ matrix models, one with coupling
$g_1 = 1/2\mu_1$ the other with coupling $g_2 = 1/2\mu_2$.
Outside of perturbation theory, large excitations with energies of
(absolute value) of order $1/g_i$ can couple the two parallel
universes. This is similar to the statement that large energies in
the usual matrix model can splash over the potential barrier.

In the last step we make a single universe with a finite number of
fermions and a finite spatial volume. A semiclassical  description
of this state, which we denote $|crescent\rangle$, is obtained at
$t=0$ by simply terminating and connecting the upper and lower
Fermi surfaces at some (large) value of the spatial coordinate
$x=x_{max}$. The filled crescent-shaped region of phase space at
$t=0$ is depicted in figure 4(a). The finite number of fermions is
given by the finite volume of phase space. A picture of the filled
energy region is given in figure 4(b). $|crescent\rangle$ is time
dependent. This region of phase space follows along classical
trajectories, sliding around the hyperbolae and then heading out
to $x=\infty$ in an elongated shape.

\fig{(a) Phase space diagram of the Fermi sea at $t=0$ for the
crescent cosmology. (b) In a position-energy diagram at $t=0$ the
potential is filled up to $x=x_{max}$.}{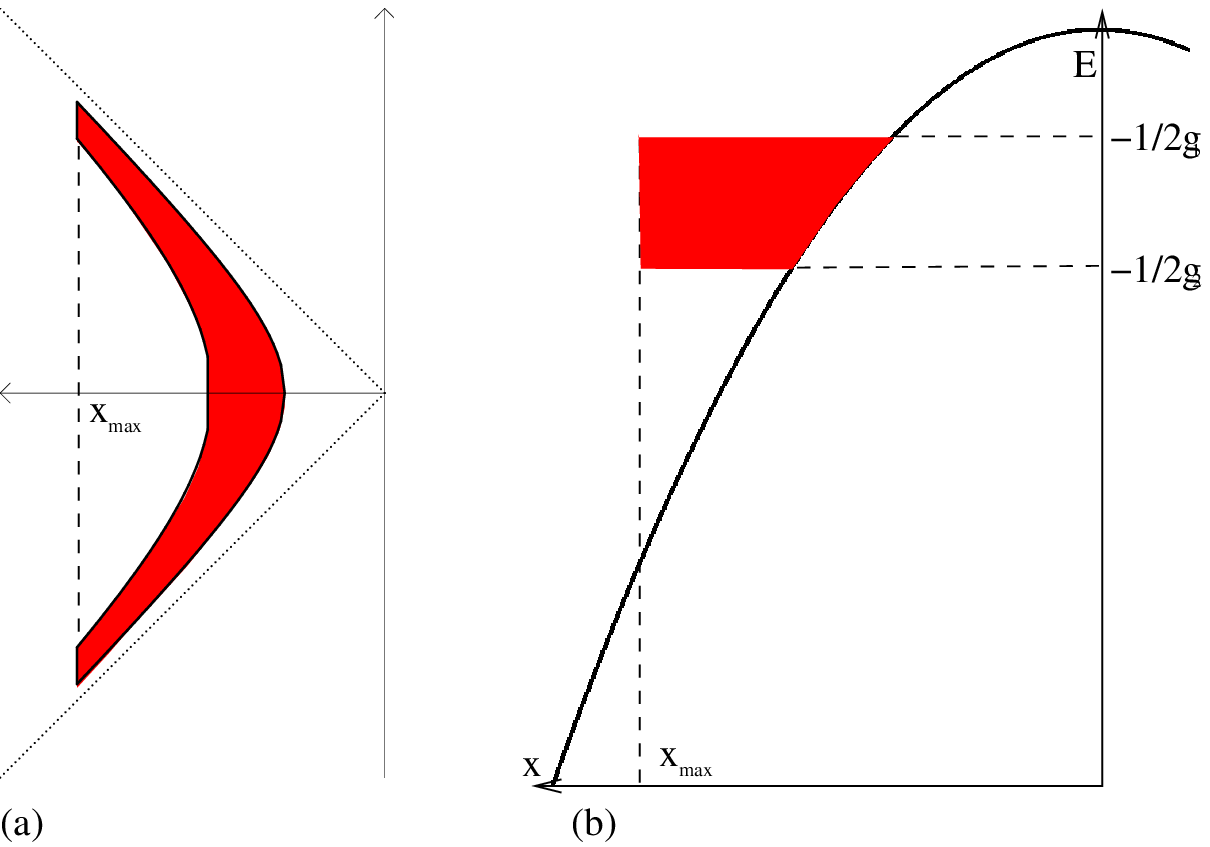}{0.0truein}


What is the physical interpretation of the state
$|crescent\rangle$? It can be found by taking $x_{max}$ to be
extremely large so that it is far out in the weakly coupled
asymptotic region. Then we can consider scattering experiments  in
which we scatter tachyons off the Liouville wall from some fixed
spatial position $x_s \ll  x_{max}$ but still deep in the weak
coupling region. Clearly such experiments will not be affected by
the termination of the Fermi sea at $x_{max}$. Therefore at,
$t=0$, we can to a good approximation describe the system as two
universes, each of which has a tachyon wall at one end and is
joined to the other at the large position $x_{max}$. The stringy
junction at which the two universes meet involves energies of
order $1\over g$ and has no perturbative description that we know
of.

As $t$ evolves forward in time, the termination points of the
Fermi sea do not remain at the same spatial position. The lower
termination point -- which we refer to as the head of the universe
-- moves away from the Liouville wall at the speed of light.  The
``tail'' of the universe initially moves toward the Liouville
wall, but then bounces off of it  towards spatial infinity. In the
far future both the head and the tail of the universe -- i.e. the
junctions between the parallel universes -- are headed away from
the Liouville wall at the speed of light. The only tachyon
excitations which survive are those which are also moving at
the speed of light sandwiched between the head and tail. The
Penrose diagram of this cosmology is depicted in figure 5.
\fig{The Penrose diagram for the crescent cosmology consists of
two null strips joined at the Liouville wall.  Each point on the
diagram corresponds to two points (one in each universe), except
for the double-lined edges, at which the two universes join.
Tachyon excitations can scatter from past to future null infinity.
} {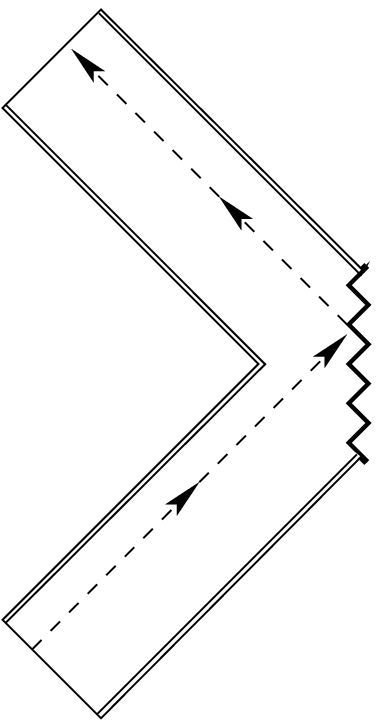}{0.0truein}

At late and early times, the universe (with the exception of the
defect region) is at weak coupling. If the tachyon wall is at
strong coupling, it bounces through a strong coupling phase to get
from one weakly coupled region to the other. In the tachyon
picture this will look like a universe which bounces through a
singularity.

An analog of the bosonization formula \ppm\ can be worked out here
and leads to two chiral, rather than one nonchiral, fields.
Consider part of a Fermi sea sandwiched between two hyperbola
defined by $x^2 - (p_i)^2 = g_i^{-1}$ where $i=1,2$ and $g_1 >
g_2$. We will be interested in the $p_i > 0$ and large $-x$
region, corresponding to the far past of the $|crescent\rangle$
state (the treatement of the far future is analogous). 
Allowing $p_i$ to fluctuate around this solution, the
equation of motion is \eqn\pEoM{
\partial_t p_i = x - (\partial_x p_i) p_i
} and the commutation relationship is \eqn\commP{ [p_i(x),
p_j(x')] = 2\pi i(-)^j \delta_{ij}\partial_x \delta(x-x')~. } Define
two new fields, $A_i(q)$, where $x=-e^{-q}$, by \eqn\ABdef{ p_i =
e^{-q} - e^q A_i ~. } The equation of motion for these two fields
is \eqn\ABEoM{ (\partial_q - \partial_t) A_i = 0 ~, } so they are
chiral (right-moving) fields.  They are linked by a commutation
relationship \eqn\commAB{ [A_i(q), A_j(q')] =  2\pi i \delta_{ij}
(-)^j\partial_q \delta(q-q')~. } These are the commutation
relation for the (derivatives of) two chiral bosons, one with
positive and one with negative norm. Thus, the bosonic
interpretation of this early-time patch of Fermi sea is that it corresponds
to two chiral, right-moving fields, consistent with the analysis
above.

\subsec{Validity of the semiclassical approximation}

The previous section described quantum states of free fermions
as regions of Fermi liquid moving under Hamiltonian flow in phase
space. Of course there are many families of exact quantum states
with any prescribed semiclassical limit. For our purposes so far
it has been important to know that there do exist exact quantum
states whose semiclassical limit has the requisite properties, but
it was unnecessary to specify which one we were considering.

However we would like to know when the semiclassical description
is valid and how or when it breaks down. Validity of the
semiclassical approximation requires that the size of the filled
region of phase space in units of $\hbar$ is large (in the
preceding sections $\hbar$ has been set to one). Equivalently the
number of fermions must obey $N \gg 1$. In addition the Fermi
surface must vary slowly on the scale set by $\hbar$. Another way
of saying this is that quantum effects produce a coarse graining
of size $\hbar$ in all the phase space diagrams.

For the ground states $|\mu\rangle$, this is equivalent to the
perhaps more familiar statement that $g_s \sim 1/\mu \ll 1$. 
The classical hamiltonian flow in  phase space is invariant under a
scaling symmetry $x \to e^\rho x, ~~p\to e^\rho p$
and $E \to e^{2\rho}E$. Therefore scaling $g_s \sim 1/E$ is the
same as scaling the size of fundamental phase space cells, and
$g_s \to 0$ is the semiclassical Fermi liquid limit.

More generally, given any trajectory of a droplet of Fermi liquid
in phase space with surface  $p(x)=f(x,t)$,  there is a family of
droplets $p(x)=e^{-\rho}f(e^\rho x,t)$ containing of order
$e^{2\rho}$ fermions. The semiclassical expansion is an expansion
in $e^{-\rho}$. It can also be thought of as an expansion in the
size of the phase space cells.

The validity of the semiclassical approximation at any moment in
time does not guarantee its eternal validity. Consider for example
the crescent cosmology of figure 4(a). In the far future, the filled
region of phase space stretches in the horizontal ($x$) direction,
while squeezing in the vertical ($p$) direction.  Depending on the
precise form of the initial quantum state, the uncertainly in $p$
could come to dominate over the semiclassical spread of $p$, and
the state would cease to behave semiclassically. While we have not
studied this question in detail, clearly the very long time
behavior of the cosmologies discussed herein could differ from the
semiclassical picture.

\newsec{General droplet cosmology}

The construction of the previous section has many generalizations.
In order to better understand what kind of universes can be
constructed from the Fermi sea filling various regions in phase
space, in this section we discuss a few aspects of the Hamiltonian
flow in the $(x,p)$ plane implied by the Hamiltonian $H = \half
(p^2 - x^2)$.

The Fermi surface moves following the equations of motion $\dot p
= x$ and $\dot x = p$. Consider a pair of points $(x_1, p_1)$ and
$(x_2, p_2)$.  The slope of the line joining them \eqn\slope{
\alpha \equiv (p_1-p_2)/(x_1-x_2) } evolves according to the
equation \eqn\SlopeEvolution{ \dot \alpha = 1 - \alpha^2~. } The
slope evolves independently of its absolute position. As a result,
straight lines stay straight and parallel ones stay parallel. The
solution is either \eqn\LargeSlope{ \alpha(t) = \coth(t-t_0) } for
$|\alpha| > 1$ or \eqn\SmallSlope{ \alpha(t) = \tanh(t-t_0) } for
$|\alpha| < 1$. Take a curve in the $x=p$ plane at some instant,
and consider its evolution.  We will divide the curve into two
regions: Steep and Shallow.  The Steep part has $|slope| > 1$ and
the Shallow $|slope| < 1$. At $t=-\infty$, the entire Steep region
has slope arbitrarily close to $-1$. At the curve evolves from
$t=-\infty$ to $t=+\infty$, the point at which it is
vertical moves through the Steep region, and eventually, at
$t=+\infty$, the entire region has slope arbitrarily close to $1$.
This is the type of region we have focused on up to this point.
The (almost) straight shape at $t=\pm\infty$ corresponds to the
flat weak coupling region in the Liouville model. Every small
perturbation about these asymptotic shapes must pass through the
vertical point on the Fermi surface, where the collective fields
are strongly coupled.  This corresponds to scattering off the
Liouville wall.  By a small perturbation we mean here one
in which the absolute value of the slope remains greater than
one, $\it{i.e.}$ a perturbation that does not create a Shallow
region within the Steep region.

The Shallow region is quite different: it's slope remains finite.
It is not clear what the space-time interpretation of the
collective coordinates describing the fluctuations about this
region is.  Classically, the Steep and the Shallow regions evolve
independently, and fluctuations do not move from one to the other.
Quantum mechanically, these region do interact, corresponding to
stringy interaction in the bosonic picture.

Consider in this light the state $|crescent\rangle$ described in
the previous section.  It contains two Steep regions, and the
physical interpretation of these was given above.  These two steep
regions must be connected by Shallow regions. The mysterious
stringy connection between the two parallel universes happens at
the Shallow regions.

One way to avoid having a Shallow region in the $|crescent\rangle$
picture is to join the two Steep regions with a cusp. The two
parallel universes are then joined through some sort of stringy
defect at the cusp.

The Shallow region need not be small compared to the Steep region.
We expect that the evolution of a fairly general Fermi sea droplet
corresponds to a closed cosmology.  Such a droplet can be divided
into four pieces: the Right and Left Steep regions, and the Top
and Bottom Shallow regions, as illustrated in figure 6. \fig{A
fairly general Fermi droplet can be divided into four regions that
behave differently under evolution.}{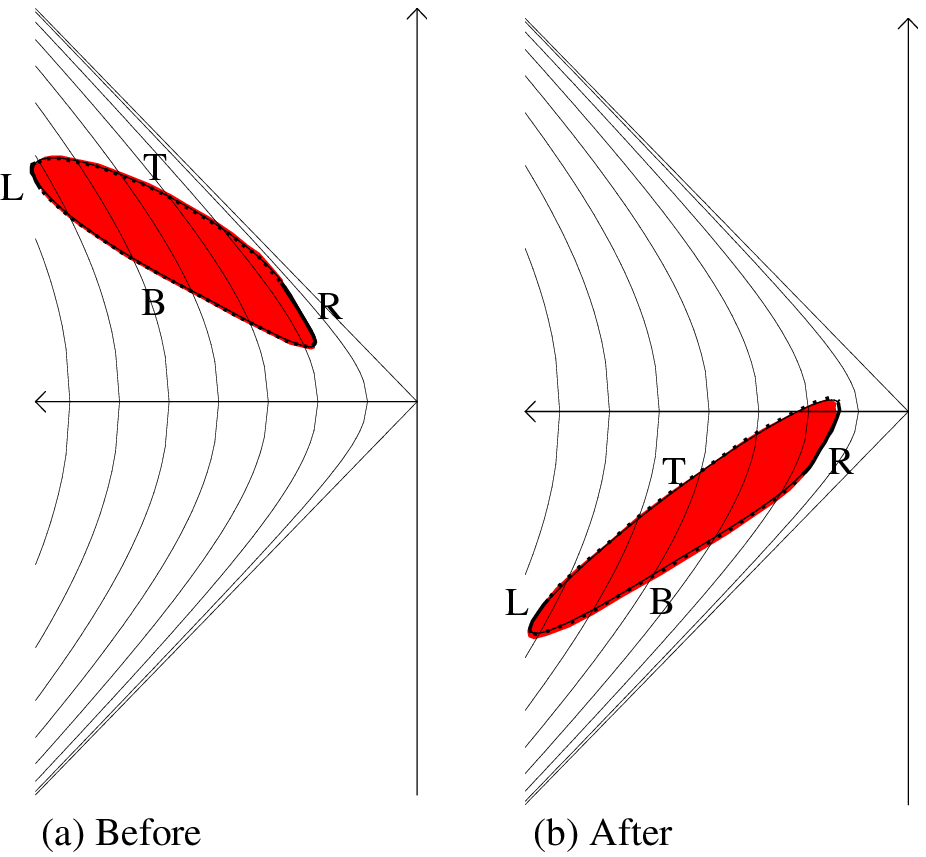}{0.0truein} The
physics in the asymptotic past/future is similar to that of the
$|crescent\rangle$ shape, though with all four regions
contributing to large portions of the universe. In the asymptotic
future, for example, the Right and the Bottom regions form a
single region, which should be described as a whole by the
standard methods used so far.

It would be interesting to discuss such a closed cosmological
universe in the same detail as we have been able to discuss the
universe with the moving Liouville wall arising from the draining
of the Fermi sea. As a step in this direction, we give a simple 
solution of the flow equations which corresponds to such a droplet
\eqn\droplet{ e^{-2t} (x+p+e^t)^2 + e^{2t} (x-p+e^{-t})^2 = a^2 ~.}
This is only one example from a large family of similar solutions.
For small enough $a$ it describes a moving and evolving ellipse
which at time $t=0$ is a circle of centered at $(x,p)=(-1,0)$. It
would be quite interesting to obtain the tachyon field
corresponding to this solution, but the technique used in this
paper does not seem to apply.

As a last comment, one might also consider solutions which extend
over the top of the Fermi sea. Trajectories in the upper quadrant
of phase space are related by a canonical transformation
$x\leftrightarrow p$ to those in the left quadrant, so these
should also have a semiclassical interpretation in some regions.

\newsec{SD-branes and higher dimensions}

The cosmologies of the preceding section can be viewed as a
"near-horizon" decoupling limit of a stack of $N$ $c=1$ sD-branes.
Indeed this project arose in an attempt to understand such a
decoupling limit for sD-branes. Viewed in this way, our results
may have a generalization to higher dimensions. In this section we
explain this connection.

    In critical string theory the
     spectrum of open strings on an unstable D-brane has a
tachyon field $T_{open}$ which lives on the boundary of the string
worldsheet. This has s-brane solutions \mgas\ which are
represented \sen\ by the boundary CFT interaction $T_{open}\sim
\lambda \cosh X^0 $. When $\lambda=\half$, this interaction
rotates Neumann to Dirichlet boundary conditions and the s-brane
becomes an array of sD-branes located on the imaginary time axis.
The sD-brane array state is perturbatively dual to a state $|C
\rangle$ which is a particular configuration of imploding and
exploding closed strings \refs{\msy ,\GaiottoRM}.

A first question to ask is: what is the $c=1$ analog, in the free
Fermi picture, of the closed string state $|C\rangle$? According
to the picture developed in \refs{\hv \MartinecKA -\kms}, in the
semiclassical limit an unstable D0-brane with $T_{open}\sim
\lambda \cosh X^0 $, is represented in the free fermion picture by
a single fermion orbiting above the Fermi sea. When
$\lambda=\half$, the fermion energy is such that it skims the
surface of the Fermi sea and can be described as a closed string
excitation. Hence the $c=1$ analog of the sD-brane array state is a
single fermion skimming the surface of the Fermi sea.

In any dimension, it is of interest to try to find a large $N$
decoupling limit of a stack of $N$ sD-branes, which may be
holographically dual to a supergravity solution. For $c=1$, taking
$N$ sD-branes corresponds to taking $N$ fermions clustered
together. This indeed has a nice semiclassical  limit which we
described above as an isolated droplet of Fermi liquid. For a
droplet orbiting above a bulk Fermi sea, it is a decoupling limit
because small fluctuations of the bulk and droplet components of
the Fermi surface do not talk to one another. The construction of
such a decoupling limit in the higher dimensional case is an
intriguing open challenge.

\centerline{\bf Acknowledgements}
 This work was supported in part by DOE grant DE-FG02-91ER40654 and the
Harvard Society of Fellows. We are grateful to P.-M. Ho,  V. Kazakov,
J. Maldacena, J. McGreevy, S.Minwalla, J. Polchinski, N. Seiberg, 
S. Shenker and T. Takayanagi for useful conversations.

\listrefs
\end